\documentclass[epsfig,11pt]{article}

\usepackage{epsfig}
\usepackage{graphicx}
\usepackage{array}
\usepackage{caption}
\usepackage{subcaption}
\usepackage{slashed}
\usepackage{amsmath}
\usepackage{braket}
\usepackage{amssymb}
\usepackage{braket}
\usepackage{comment}

\newcommand{\beq}{\begin{equation}}   
\newcommand{\eeq}{\end{equation}}
\newcommand{\beqn}{\begin{eqnarray}}   
\newcommand{\eeqn}{\end{eqnarray}}

\newcommand{\bea}{\begin{eqnarray}}
\newcommand{\eea}{\end{eqnarray}}
\newcommand{\be}{\begin{equation}}
\newcommand{\ee}{\end{equation}}
\newcommand{\bead}{\begin{aligned}}
\newcommand{\eead}{\end{aligned}}

\newcommand{\gsim}{\lower.7ex\hbox{$
\;\stackrel{\textstyle>}{\sim}\;$}}
\newcommand{\lsim}{\lower.7ex\hbox{$
\;\stackrel{\textstyle<}{\sim}\;$}}
\begin{document}

\begin{titlepage}

\begin{flushright}
FTPI-MINN-15/46\\
UMN-TH-3509/15
\end{flushright}

\vspace{0.7cm}

\begin{center}
{ \Large \bf  More on Two-Dimensional \boldmath{$O(N)$} Models\\[2mm] with $\mathcal{N} = (0,1)$ 
Supersymmetry }
\end{center}
\vspace{0.6cm}

\begin{center}
 {\large 
Adam Peterson$^a$, Evgeniy Kurianovych$^a$, and Mikhail Shifman$^{a,b}$}
\end {center}

\vspace{3mm}
 
\begin{center}

$^a${\em University of Minnesota School of Physics and Astronomy,
Minneapolis, MN 55455, USA}
\\[1mm]
$^b${\em William I. Fine Theoretical Physics Institute, University of Minnesota,
Minneapolis, MN 55455, USA}\\[1mm]

\end {center}

\vspace{2cm}

\begin{center}
{\large\bf Abstract}
\end{center}

We study the behavior of two dimensional supersymmetric connections of $n$ copies of $O(N)$ models with an $\mathcal{N} = (0,1)$ heterotic deformation generated by a right moving fermion.  We develop the model in analogy with the connected $\mathcal{N}=(0,2)$ $CP(N-1)$ models for the case of a single connecting fermionic superfield.  We calculate the effective potential in the large $N$ limit and determine the vacuum field configurations.  Similarily to other SUSY connected models we find that SUSY is unbroken under certain conditions despite the vanishing of the Witten index.  Specifically, this preservation of SUSY occurs when we have an even number $n$ of $O(N)$ families.  As in previous cases we show that this result follows from a $Z_n$ symmetry under a particular exchange of the $O(N)$ families.  This leads to a definition of a modified Witten index, which gaurantees the preservation of SUSY in this case.

\hspace{0.3cm}

\end{titlepage}
%%%%%%%%%%%%

%%%%%%%%%%%%%%%%%%%%%%%%%%%%%%%%
%%%%%%%new definitions %%%%%%%%%%%%%%%%
%%%%%%%%%%%%%%%%%%%%%%%%%%%%%%%%

\section{Introduction}
\label{intro}

Two-dimensional chiral sigma models are known for a long time (see e.g. \cite{1}). Some recent works devoted to such models are 
 \cite{Witten:2005px,bai2,Yagi:2010tp,Adams:2003zy,Melnikov:2012hk,Jia,Gadde:2013lxa,Gadde:2014ppa,Cui:2011rz, CCSV1,Cui:2010si,Cui:2011uw,Shifman:2008kj,Shifman:2014lva,16}.
A revival of interest is due to the fact that 
 chiral $\mathcal{N}\!=(0,2)$ sigma models
emerged as low-energy world sheet theories on non-Abelian strings supported in some $\mathcal{N}\!\!=\!1$
four-dimensional Yang-Mills theories \cite{Edalati:2007vk} (for a review see  \cite{Shifman:2014jba}). 

In this paper we will consider two questions. First, we will consider a non-minimal  $\mathcal{N} = (0,1)$  $O(N)$ model and construct ``connected copies," following the example of Ref. \cite{Shifman:2014lva} that
addressed $\mathcal{N} = (0,2)$  $CP(N-1)$ models. As in \cite{Shifman:2014lva}, we prove that (a) the spontaneous breaking of supersymmetry disappears in the $1/N$ solution. Moreover, we introduce a generalized Witten index suitable for our model and show that it does not vanish (the conventional Witten index vanishes). 
Second, we construct the large-$N$ solution of the minimal  $\mathcal{N} = (0,1)$  $O(N)$ model. Note that the minimal chiral $O(N)$ models are free from anomalies and thus selfconsistent for any $N$ \cite{Moore:1984ws,16}.
Nonminimal models are free from anomalies by construction.

Connecting $n$ copies of $O(N)$ sigma models will be performed in the manner described below, following the pattern of \cite{Shifman:2014lva}. This will result in the additional $Z_n$ symmetry under exchanges of the $O(N)$ sectors, which, in turn,  is responsible for supersymmtry restotration and for the existence of 
a modified Witten index $I_P\neq 0$. The latter guarantees that the restoration of supersymmetry observed to the 
leading order in $1/N$ is in fact exact.

The $O(N)$ models with real target spaces $S^{(N-1)}$  and $\mathcal{N} = (0,1)$ supersymmetry have their peculiarities  \cite{Sakamoto:1984zk}. In particular, to our knowledge, unlike the $CP(N-1)$ case, the $O(N)$ models do not follow from bulk four-dimensional Yang-Mills solitons.  They are found more frequently in effective field theories in condensed matter systems (see, for example, \cite{Volovik:2009}).  Additionally, many convenient simplifications following from the chiral behavior of K\"ahler manifolds \cite{Zumino:1979et} will be absent in the $O(N)$ case.  Regardless, we will find that under certain constraints, many results from the connected $CP(N-1)$ models will carry over to the $O(N)$ case.  In particular, we show that these models have vanishing Witten index and a vanishing vacuum energy at one loop and thus unbroken supersymmetry (see results in \cite{Alvarez:1977qs}).  Additionally, just as in the $CP(N-1)$ case, this result follows from the $Z_n$ symmetry of exchanges between the $n$ $O(N)$ sectors.  Using the exchange symmetry between $O(N)$ families a modified Witten index may be defined for the $O(N)$ case as well.  

Large-$N$ solution of the $O(N)$ sigma model with ${\cal N}= (1,1)$ supersymmetry was constructed in \cite{Alvarez:1977qs}, while its generaization to nonminimal ${\cal N}= (0,1)$ models was presented in \cite{Koroteev:2010gt}.

The paper is organized as follows.  We will begin with a discussion of generalities for $\mathcal{N} = (0,1)$ models in two dimensions.  In the following section we will develope the minimal $O(N)$ and non-minimal connected $O(N)^n$ model with the $\mathcal{N}=(0,1)$ deformation generated by the connecting fermion superfield.  For both cases will calculate the effective potential and determine the vacua.  We will then explore the conditions for SUSY preservation.  For the non-minimal models this will be accomplished by determining the mass spectrum and calculating a modified Witten index, which we will develop using the $Z_n$ symmetry of our model.  We will conclude with a comparison of the $O(N)$ and $CP(N-1)$ cases.

\section{Generalities}
The $\mathcal{N} = (1,1)$ theories in two dimensions have two real supercharges which can be defined in the Majorana-Weyl bases as $Q_L$ and $Q_R$ with the defining anti-commutator
\begin{equation}
\left\{Q_\alpha,Q_\beta\right\} = 2P_\mu(\gamma^\mu)_{\alpha\beta}=2\left(\begin{array}{cc}
E-P & 0\\
0 & E+P\\
\end{array}\right)_{\alpha\beta}.
\end{equation}
In the following we will be considering $\mathcal{N} = (0,1)$ theories with only the single $Q_L$ supercharge as discussed in \cite{Sakamoto:1984zk}.  It will thus be convenient to decompose the $\mathcal{N} = (1,1)$ superfields in terms of $\mathcal{N} = (0,1)$ fields.  The $\mathcal{N} = (1,1)$ superfields my be written as
\begin{equation}
\Phi(x,\theta_L,\theta_R) = \phi +\bar{\theta}\psi + \frac{1}{2}\bar{\theta}\theta F = A(x,\theta_R)-{\rm i}\theta_L B(x,\theta_R)
\end{equation}
where $A(x,\theta_R)$ is a scalar $\mathcal{N}=(0,1)$ superfield and $B(x,\theta_R)$ is a right-moving fermionic $\mathcal{N}=(0,1)$ superfield.  More explicitly:
\begin{align}
&A(x,\theta_R) = \phi(x) +{\rm i} \theta_R\psi_L(x) \nonumber \\
&B(x,\theta_R) = \psi_R(x) +\theta_R F(x).
\end{align}

We may now use the SUSY transformations of a superfield (shown in the appendix) to determine how the $\mathcal{N} = (0,1)$ fields transform.  It is easy to show that for $\theta_R \rightarrow \theta_R +\epsilon_R$
\begin{align}
&\delta \phi = {\rm i } \epsilon_R \psi_L\,, \nonumber \\[1mm]
&\delta \psi_L = -2\epsilon_R \partial_L \phi\,, \nonumber \\[1mm]
&\delta \psi_R = \epsilon_R F\,, \nonumber \\[1mm]
&\delta F = -2{\rm i} \epsilon_R \partial_L \psi_R\,.
\end{align}
It is clear that both superfields $A(x,\theta_R)$ and $B(x,\theta_R)$ are irreducible under $Q_L$.

Appropriate Lagrangians for $\mathcal{N} = (0,1)$ models may be constructed by integration over the single Grassmann coordinate $\theta_R$ provided they are Lorentz invariant.  The kinetic terms for $A(x,\theta)$ and $B(x,\theta)$ may be written as
\begin{align}
&\mathcal{L}_{{\rm kin},A}=-\int d\theta \left(\partial_R A\right)\left( D_L A\right), \nonumber \\[1mm]
&\mathcal{L}_{{\rm kin},B}=\frac{{\rm i}}{2}\int d\theta\, B D_L B\,.
\end{align}
Here the supercovariant derivative is written as
\begin{equation}
D_L = -{\rm i}\frac{\partial}{\partial\theta_R}-2\theta_R\partial_L\,.
\end{equation}

Non-kinetic terms for $A(x,\theta)$ and $B(x,\theta)$ may also be defined provided that the integrand for $\theta$ integration is a right-moving fermion superfield.  We will make use of these requirements below.

In the sections below we will consider connected models of the type $O(N)^n$ where $n$ individual $O(N)$ sectors are connected by a right moving fermion fields transforming trivially under $O(N)$ rotations.  In general, connected-sector models are constructed for arbitrary $n$ as illustrated in Figure 1. 

\begin{figure}[ptb]
\centering
\includegraphics[width=0.6\linewidth]{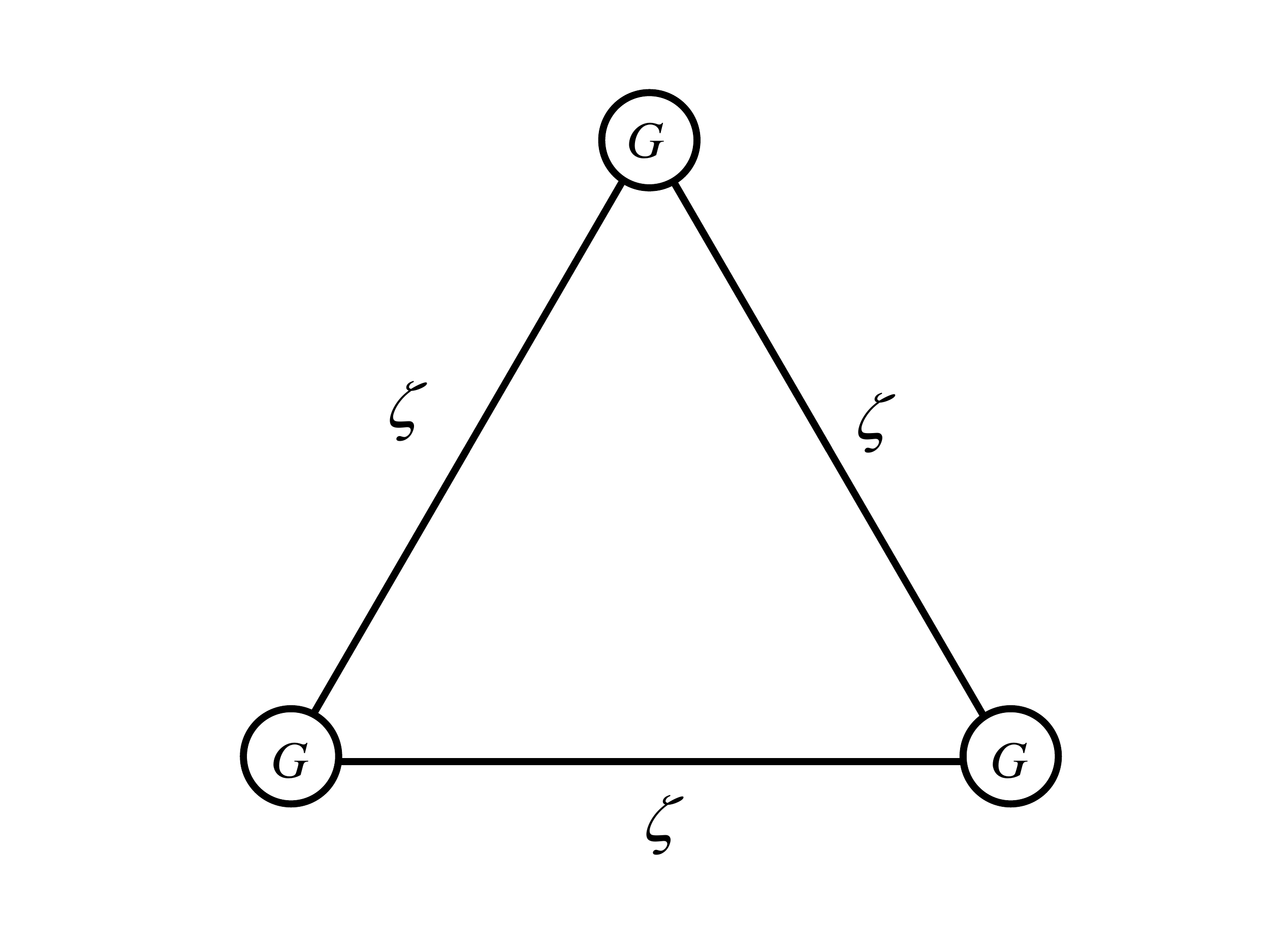}
\caption{The connected diagram above illustrates the mechanism by which different $O(N)$ sectors (represented by $G$ above) are connected by fermion superfields $\zeta$ which do not transform under $O(N)$.}%
\end{figure}

\section{Models}

We will discuss two versions of the deformed $O(N)$ model in this section.  The first case we consider will be the minimal $\mathcal{N} = (0,1)$ $O(N)$ model with only left moving fermions $\psi^i_L$ appearing in the Lagrangian.  

\subsection{The minimal model}

For completeness we begin with a short discussion of the minimal $\mathcal{N} = (0,1)$ $O(N)$ model where the deformation is achieved simply by throwing out the right-moving fermions in the superpotential.  In many cases this procedure leads to anomalies due to the target space manifold structure  \cite{Moore:1984ws}.  However, the $O(N)$ model is free from these anomalies.

We define the superfields
\begin{align}
&N^i = n^i + {\rm i}\theta_R \psi^i_L\,, \nonumber \\[1mm]
&\Lambda_R = \lambda_R +\theta_R D\,.
\end{align}
Here $\Lambda$ is an auxillary superfield introduced to constrain the $O(N)$ field $N^i$.  The Lagrangian for the minimal $O(N)$ model may be written as
\begin{equation}
\mathcal{L} = \frac{1}{2}\int d\theta \left[ -2 \left(\partial_R N^i \right)\left( D_L N^i\right) - \Lambda_R 
\left(N^i N^i -\frac{1}{g^2}\right)\right].
\label{MinimalModel}
\end{equation}
In components the Lagrangian reads
\begin{equation}
\mathcal{L} = \frac{1}{2}\partial_\mu n^i \partial^\mu n^i+{\rm i} \psi^i_L \partial_R \psi^i_L+{\rm i}\lambda_R \psi^i_L n^i-\frac{1}{2}D\left(n^i n^i -\frac{1}{g^2}\right).
\label{MinimalModelComp}
\end{equation}
For the purpose of determining vacuum field configurations it is most convenient to integrate over the fields $n^i$ setting $\lambda_R = 0$ as required by Lorentz symmetry.  Performing this calculation (see \cite{Witten:1978bc}) we find the prefactor
\begin{equation}
\frac{1}{{\rm Det}\left(\partial^2+D\right)^{N/2}}\,.
\end{equation}
We may then write the effective action ignoring $D$-independent terms from integration over $\psi^i_L$,
\begin{equation}
\Gamma = \frac{{\rm i}N}{2}{\rm Tr}\log{(\partial^2+D)}+\frac{1}{2g^2}\int d^2xD.
\end{equation}
Assuming a constant value of $D$ in the vacuum we may write the effective potential as follows:
\begin{equation}
V_{\rm eff} = \frac{N}{8\pi}\left\{ D\left(\log{\frac{M_{\rm uv}^2}{D}}+1\right)-\frac{4\pi D}{Ng^2}\right\}=\frac{N}{8\pi}D\left( \log{\frac{\Lambda^2}{D}}+1\right),
\label{MinimalVeff}
\end{equation}
where we have defined a scale parameter
\begin{equation}
\Lambda^2 = M^2_{{\rm uv}}{\rm e}^{-\frac{4\pi}{Ng^2}}.
\end{equation}
Minimizing the effective potential (\ref{MinimalVeff}) we find
\begin{equation}
\langle D \rangle = \Lambda^2.
\label{MinimalVacD}
\end{equation}
Clearly the non-zero vacuum expectation value of $\langle D \rangle$ breaks the $\mathcal{N}=(0,1)$ 
supersymmetry.\footnote{This large-$N$ result has no direct parallel in minimal (0,2) $CP(N-1)$ models 
since such a model exists only for $N=2$, see \cite{Moore:1984ws,16}. However, it is similar to the
result obtained in a {\em nonminimal} (0,2) $CP(N-1)$ models in \cite{Shifman:2008kj}.}  We will verify this result below by considering the mass spectra.

\subsection{Non-minimal (0,1)  connected \boldmath{$O(N)$} models}
\label{nmm}

For non-minimal models a similar procedure may be carried out.\footnote{The heterotic deformation of the $O(N)$ model was first considered in \cite{Koroteev:2010gt} (see also \cite{Shifman:2008kj}),  with a single SU$O(N)$ superfield $\mathcal{N}^i$ and one $\mathcal{N} = (0,1)$ fermion superfield deformation. }  We define the following superfields ($F$ is the ``flavor" index of $n$ sectors, $F=1,2, ... , n$)
\begin{align}
&\mathcal{N}^i_F = N^i_F-{\rm i}\theta_L\Psi^i_F=n^i_F+\bar{\theta}\psi^i_F+\frac{1}{2}\bar{\theta}\theta F^i_F\,,\nonumber \\[2mm]
&\mathcal{S}_F= S_F-{\rm i}\theta_L\Lambda_F=\sigma_F+\bar{\theta}\lambda_F+\frac{1}{2}\bar{\theta}\theta D_F \,,\nonumber \\[2mm]
&\mathcal{B}=-{\rm i}\theta_L B=\bar{\theta}\zeta_R+\frac{1}{2}\bar{\theta}\theta G\, , \qquad \zeta_R = \left(\begin{array}{c}
0 \\
\zeta_R
\end{array}\right),
\end{align}
where the first equality is the $\mathcal{N}=(0,1)$ decomposition of the superfields.  The $\mathcal{N}_F^i$ field represents the superfield living on $O(N)^n$ manifolds.  The auxillary fields $\mathcal{S}_F$ will provide the constraints for the $O(N)^n$ fields.  The $\mathcal{N} = (0,1)$ field $\zeta_R$ will connect the different flavours $F$ of the $O(N)^n$ fields.

The Lagrangian for our model can be written in terms of the $\mathcal{N} = (0,1)$ fields:
\begin{align}
\mathcal{L} = \frac{1}{2}\sum_{F=1}^n\int d\theta & \left[-2\partial_R N^i_F D_L N^i_F +{\rm i}Z\Psi^i_{F} D_L \Psi^i_{F}-\sqrt{Z}S_F N^i_F\Psi^i_{F} \phantom{\frac{1}{2}} \right. \nonumber \\[2mm]
&\left.-\Lambda_F\left(N^i_F N^i_F -\frac{1}{g^2}\right)-2\sqrt{\mathcal{Z}}\frac{\kappa}{g^2} S_F B\right]+\frac{{\rm i}}{2}\int d\theta \mathcal{Z} B D_L B.
\label{Lagrangian}
\end{align}
The factors $Z$ and $\mathcal{Z}$ are the field strength renomalization factors of the superfields $\Psi^i_F$ and $B$ respectively.  Here we choose $\kappa$ to scale at large $N$ just as in the $CP(N-1)$ cases which follows from the effective two-dimensional dynamics:
\begin{equation}
\kappa \sim \frac{1}{\sqrt{N}}.
\end{equation}
In components this Lagrangian reads
\begin{align}
\mathcal{L} = \sum_{F=1}^n &\left\{\frac{1}{2} \partial_\mu n^i_F \partial^\mu n^i_F +{\rm i}\psi^i_{FL}\partial_R \psi^i_{FL}+{\rm i}Z\psi^i_{FR}\partial_L \psi^i_{FR} \right. \nonumber \\[2mm]
&\;\; +\left. {\rm i}\lambda_{FR}\psi^i_{FL}n^i-{\rm i}\sqrt{Z}\lambda_{FL}\psi^i_{FR}n^i+{\rm i}\frac{\kappa}{g^2}\sqrt{\mathcal{Z}}\zeta_R\lambda_{FL} \right. \nonumber \\[2mm]
&\;\; -\left. \frac{1}{2}\left(D_F + \sigma_F^2\right)n^i_F n^i_F+{\rm i}\sqrt{Z}\sigma_F \psi^i_{FR} \psi^i_{FL}  +\frac{1}{2g^2}D_F \right\}  \nonumber \\[2mm]
& \;\; +{\rm i}\mathcal{Z}\zeta_R\partial_L\zeta_R-\frac{\kappa^2}{2g^4}\left(\sum_{F=1}^n \sigma_F\right)^2.
\label{LagrangianComp}
\end{align}
To find the effective potential we integrate over the $n^i_F$ and $\psi^i_F$ fields assuming a Lorentz invariant vacuum where $\lambda_F = 0$ while $D_F$ and $\sigma_F$ are space-time constants.  Proceding along these lines we find the prefactor
\begin{equation}
\prod_{F=1}^n \frac{{\rm Det}\left(\partial^2 + \sigma^2\right)^{N/2}}{{\rm Det}\left(\partial^2 + \sigma^2 + D\right)^{N/2}}.
\end{equation}
This leads to the one-loop correction (i.e. the leading $1/N$ term) to the potential
\begin{align}
V_{\mbox{1-loop}}= \frac{N}{8\pi}\sum_{F=1}^n \left[(D_F+\sigma_F^2)\left(\log\frac{M_{\rm uv}^2}{D_F+\sigma_F^2}+1\right) - \sigma_F^2\left(\log\frac{M_{\rm uv}^2}{\sigma_F^2}+1\right)\right].
\end{align}
Adding the one-loop correction to the potential we arrive at the expression for the effective potential
\begin{align}
V_{\rm eff} =& \frac{N}{8\pi}\left\{ \sum_{F=1}^n\left[(D_F+\sigma_F^2) \left( \log\frac{\Lambda^2}{D_F+\sigma_F^2}+1\right)+\sigma_F^2 \log \frac{\sigma_F^2}{D_F + \sigma_F^2} \right] \right.\nonumber \\
&+\left. u \left(\sum_{F=1}^n \sigma_F\right)^2\right\},
\label{EffectivePotential}
\end{align}
where
\begin{equation}
\Lambda^2 = M_{\rm uv}^2\; {\rm e}^{-\frac{4\pi}{Ng^2}}
\end{equation}
is a scaling parameter, and we have defined
\begin{equation}
u = \frac{4 \pi \kappa^2}{Ng^4}.
\end{equation}
With the $N$ counting behavior of $g$ and $\kappa$ we see that $u$ does not scale with $N$ at large $N$.
%\marginpar{\tiny Pls, discuss the $N$ dependence of $u$}

To find the ground state we first minimize (\ref{EffectivePotential}) with respect to $D_F$ to find the expression for the potential as a function of $\sigma_F$
\begin{equation}
V_{\rm eff}(\sigma_F) = \frac{N}{8\pi}\left\{ \sum_{F=1}^n \left[ \Lambda^2 + \sigma_F^2 \left( \log \frac{\sigma_F^2}{\Lambda^2}-1\right)\right] +u\left(\sum_{F=1}^n \sigma_F\right)^2\right\},
\label{EffPotSigma}
\end{equation}
where the minimization condition is satisfied at
\begin{equation}
D_F = \Lambda^2 - \sigma_F^2.
\end{equation}

We first consider the example of a single sector  $n=1$ which was  analyzed in \cite{Koroteev:2010gt}.  In this case 
\begin{equation}
\langle \sigma \rangle = \pm \Lambda {\rm e}^{-u/2}\,, \qquad \langle D \rangle = \Lambda^2 -\sigma^2\,,
\end{equation}
and the vacuum energy density is
\begin{equation}
\mathcal{E}_{\rm vac} = \frac{N}{8\pi} \Lambda^2(1- {\rm e}^{-u}).
\end{equation}
Clearly for $u \neq 0$ supersymmetry is completely broken in the vacuum.  These and additional details can be found in \cite{Koroteev:2010gt}.

We may extend this analysis for general values of $n$.  Both terms in (\ref{EffPotSigma}) are semi-positive-definite.  Thus we see that for $V_{\rm eff}$ to vanish both terms in (\ref{EffPotSigma}) must vanish separately.  For odd $n$ the second term in (\ref{EffPotSigma}) is positive and thus supersymmetry is necessarily broken.  On the other hand for even values of $n$ the potential is vanishing at
\begin{equation}
\langle \sigma_F \rangle = \pm \Lambda, \mbox{ and } \langle D_F \rangle = 0\,,
\end{equation}
where the positive sign is chosen for half of the $\sigma_F$ fields and the negative sign for the remaining half.  Thus supersymmetry appears to be unbroken if $n$ is even. The above consideration proves supersymmetry restoration in the leading order in $1/N$. Below we will see that this is an exact statement, fully equivalent to that of \cite{Shifman:2014lva}.

\section{Effective Lagrangian and mass spectrum}
We begin by discussing the mass spectrum of the minimal model considered above.  From the vacuum expectation value of $D$ (\ref{MinimalVacD}) determined at one loop we see that the $n^i$ and $\psi^i_L$ fields have the masses
\begin{equation}
m_n = \Lambda, \;\; m_\psi = 0.
\end{equation}
We may additionally expand the effective Lagrangian for the auxillary $\lambda_R$ and $D$ following from the loop diagrams in Figure 2 giving:
\begin{equation}
\mathcal{L}_{\rm eff} = \frac{1}{e_\lambda^2}\lambda_R \partial_L \lambda_R + (\mbox{$D$ kinetic term}),
\label{MinimalEffLagrangian}
\end{equation}
where
\begin{equation}
\frac{1}{e_\lambda^2} =\frac{N}{4\pi \Lambda^2}.
\end{equation}
Here the $D$ kinetic term can be calculated from the one-loop $D$ propagator.  The $D$ propagator (denoted $D^{(D)}(p)$) may be written as:
\begin{equation}
D^{(D)}(p) = -\frac{2}{\Gamma(p)},
\end{equation}
where
\begin{equation}
\Gamma(p) = (-{\rm i})^2\int \frac{d^2q}{(2 \pi)^2}\frac{\rm i}{q^2-m_n^2}\frac{\rm i}{(p+q)^2-m_n^2}.
\label{Dinversepropagator}
\end{equation}
It is straightforward to show that $D^{(D)}(p)$ has no poles, but only a branch cut at $p^2 = 4m_n^2$.  Thus, the $D$ field creates resonance-like states and no real particle states.  We can thus ignore the $D$ contribution for the particle spectrum of the minimal model \cite{Shifman:2012zz}.
\begin{figure}[ptb]
\centering
\includegraphics[width=0.8\linewidth]{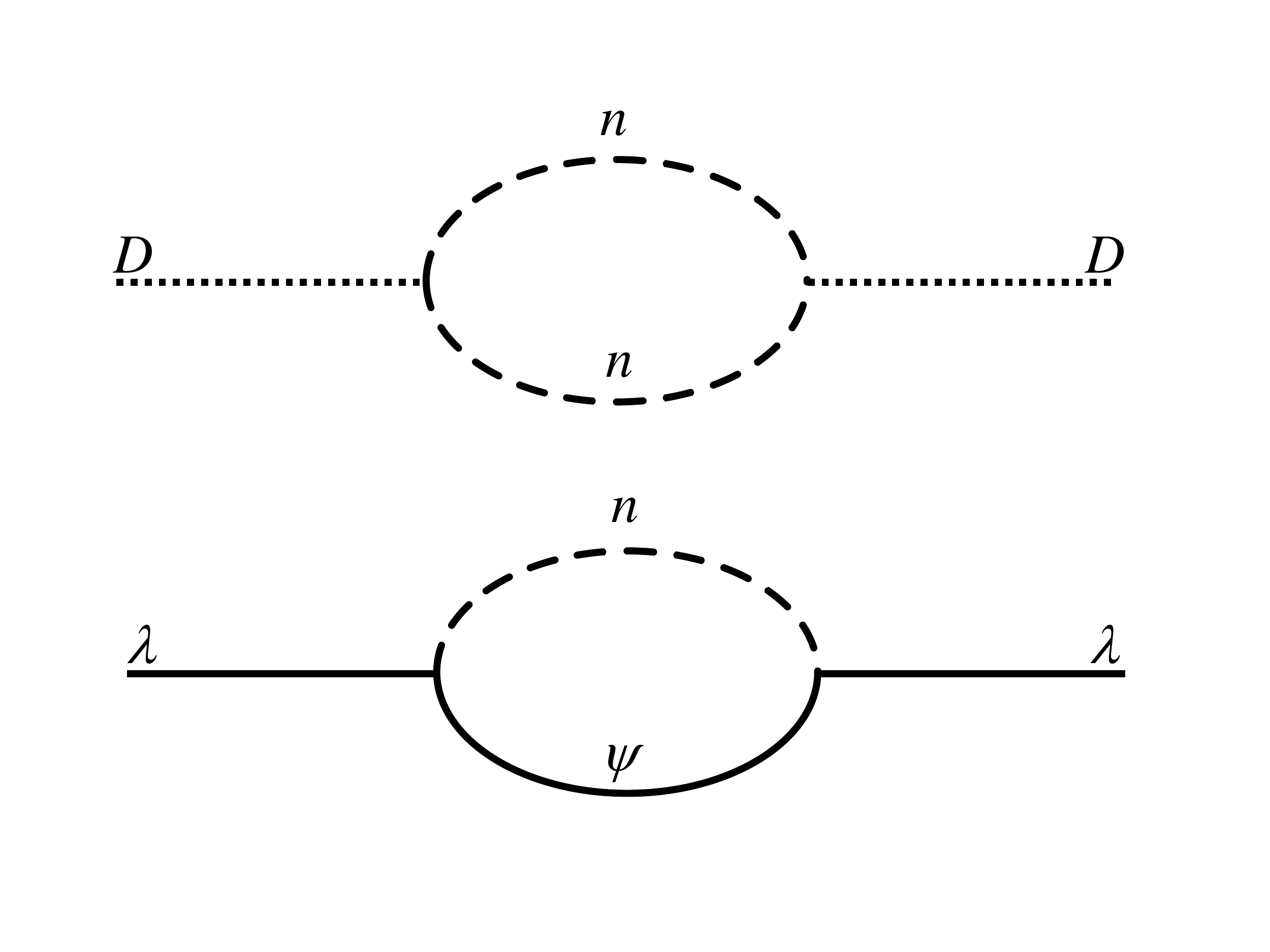}
\caption{The two diagrams above show the one-loop contribution to the $D$ (upper) and $\lambda$ (lower) propagators leading to dynamically generated kinetic terms for the auxillary fields.}%
\end{figure}

Returning to the effective Lagragian (\ref{MinimalEffLagrangian}) we see that $\lambda_R$ creates asymptotic particle states of zero mass:
\begin{equation}
m_\lambda = 0.
\end{equation}
Thus $\lambda_R$ is the Goldstino indicating SUSY breaking for this model.

More parallels with the $CP(N-1)^n$ case can be observed when one considers excitations (particles) in the $O(N)^n$ models.  The lowest excitations can be determined from the mass eigenstates of the effective Lagrangian.  For this purpose the kinetic terms for $\sigma_F$, and $\lambda_F$ as well as their interactions must be calculated to one-loop order.  This method is exact in the large-$N$ limit.

It is useful to first consider the mass spectrum of the $n=1$ model as discussed in \cite{Koroteev:2010gt}.  Considering the loop diagrams shown in Figures 3 and 4, the effective Lagragnian after integration over $n^i$ and $\psi^i$ at large $N$ is calculated to be
\begin{equation}
\mathcal{L} = \zeta_R {\rm i} \partial_L \zeta_R +\frac{1}{2e^2_\sigma} \partial^\mu \sigma \partial_\mu \sigma+\frac{{\rm i }}{2e_\lambda^2}\bar{\lambda}\gamma^\mu \partial_\mu \lambda-V_{\rm eff}(\sigma^2)+\frac{1}{2}\Gamma \sigma \bar{\lambda}\lambda+{\rm i }\frac{\kappa}{g^2} \zeta_R \lambda_L\,.
\end{equation}
The coefficients of the kinetic terms can be calculated for low momentum from Figure 3 giving
\begin{align}
&\frac{1}{e_\sigma^2} = \frac{N}{24\pi}\frac{{\rm e}^u}{\Lambda^2}(1+{\rm e}^{-2u})\,, \nonumber \\[2mm]
&\frac{1}{e_\lambda^2} = \frac{N}{4\pi}\frac{1}{\Lambda^2}\frac{1-{\rm e}^{-u}(1+u)}{(1-{\rm e}^{-u})^2}\,.
\end{align}
The coefficient of the $\bar{\lambda}\lambda \sigma$ interaction can also be calculated to one loop (Figure 4) at large $N$:
\begin{equation}
\Gamma = \frac{N}{4\pi\Lambda^2}\frac{u}{1-{\rm e}^{-u}}\,.
\label{IntGamma}
\end{equation}
\begin{figure}[ptb]
\centering
\includegraphics[width=0.8\linewidth]{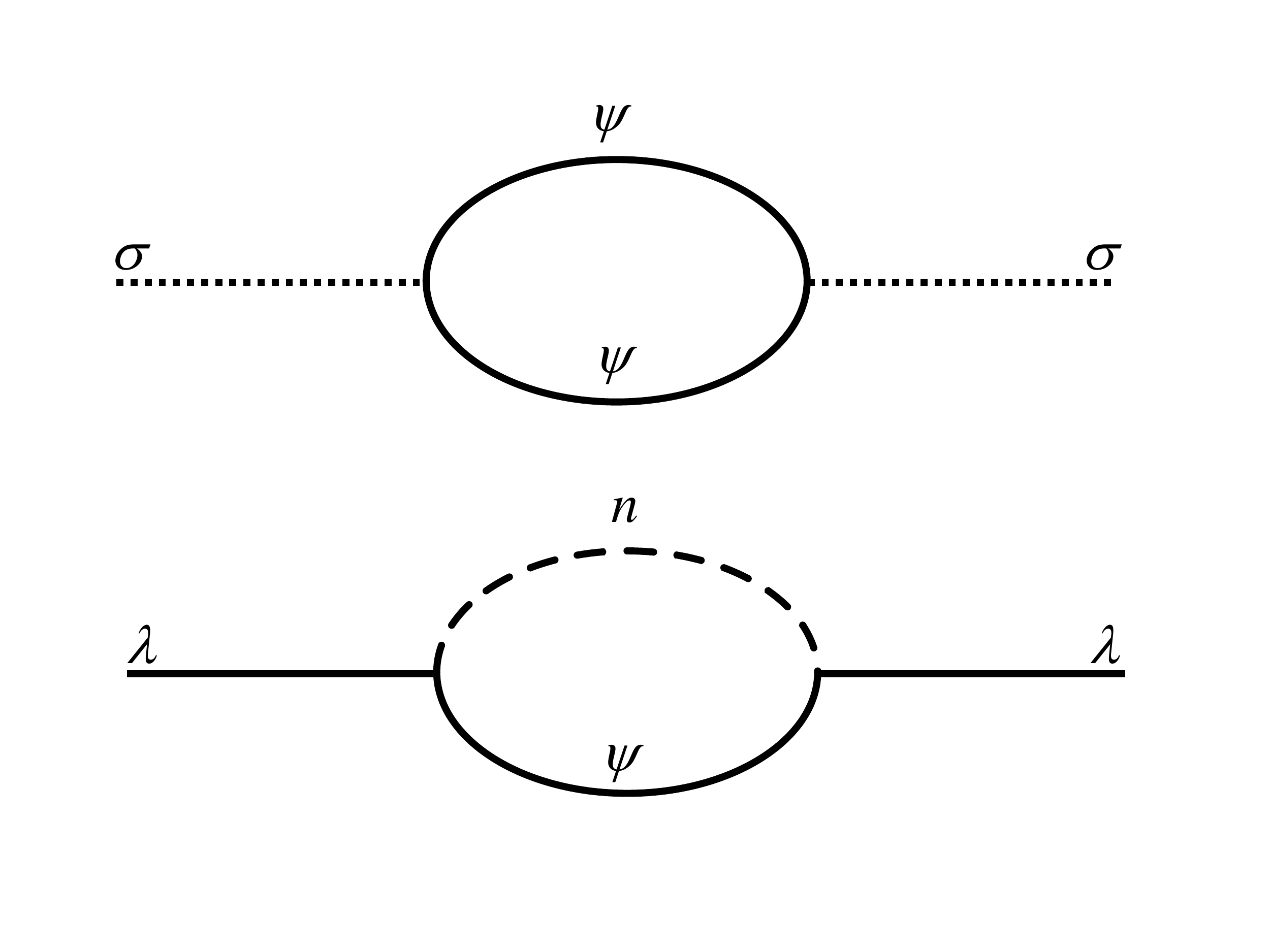}
\caption{The two diagrams above show the one-loop contribution to the $\sigma$ (upper) and $\lambda$ (lower) propagators leading to dynamically generated kinetic terms for the auxillary fields.}%
\end{figure}
\begin{figure}[ptb]
\centering
\includegraphics[width=0.8\linewidth]{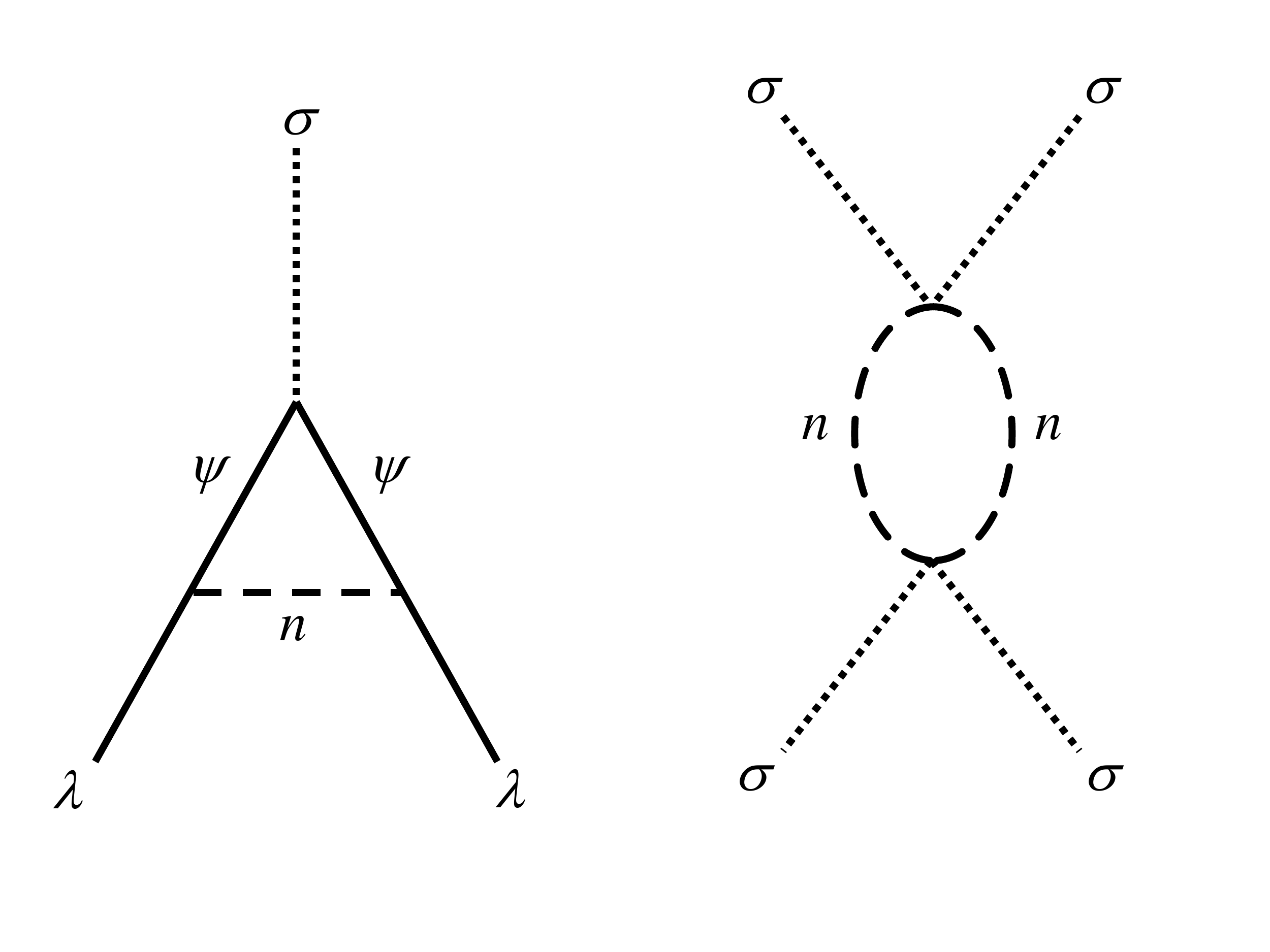}
\caption{The two diagrams above show the dynamically generated interaction vertices for the $\sigma$ and $\lambda$ fields.  The diagrams may be used to calculate the coefficient $\Gamma$ in (\ref{IntGamma}).}%
\end{figure}

It is then possible to determine the mass eigenvalues of the excitations.  For the boson mass we have
\begin{equation}
m_\sigma = \Lambda \sqrt{6} \frac{{\rm e}^{u/2}}{\sqrt{1+\frac{1}{2}{\rm e}^{2u}}}\, .
\end{equation}
One may diagonalize the fermion mass matrix leading to one massless fermion and one massive Majorana fermion
\begin{equation}
m_f = 2\Lambda\frac{\sqrt{u({\rm e}^{-u}-1)}}{{\rm e}^u-1-u}\sinh{\frac{u}{2}}\,.
\end{equation}
For non-zero $u$ the fermion and boson fields do not have equal masses and it is clear that supersymmetry is indeed spontaneously broken.  However, it is easily seen that for $u \rightarrow 0$ the kinetic coefficients and masses become equal, as expected for $\mathcal{N} =(1,1)$ restoration in this limit. The above-mentioned massless fermion is the Goldstino.

Having reviewed the $n=1$ case we may now consider the connected models.  For even values of $n$ the effective Lagrangian can be calculated to one-loop order:
\begin{align}
\mathcal{L}_{\rm eff} = \zeta_R {\rm i}\partial_L \zeta_R+&\frac{N}{8\pi}\sum_{F=1}^n \left\{\frac{1}{\Lambda^2}\left[ \frac{1}{2}(\partial_\mu \sigma_F)^2 +\lambda_{FL}{\rm i}\partial_R \lambda_{FL}+\lambda_{FR}{\rm i}\partial_L \lambda_{FR}  \phantom{\sqrt{\frac{4\pi u}{N}}}\right. \right. \nonumber \\
&\left. \phantom{\frac{1}{2}}+2{\rm i}\sigma_F \lambda_{FR}\lambda_{FL} \right]-\left( \Lambda^2 - \sigma_F^2 + \sigma_F^2 \log{ \frac{\sigma_F^2}{\Lambda^2}}\right) \nonumber \\
&\;\;\;\, \left. + 2 \sqrt{\frac{4\pi u}{N}} {\rm i} \zeta_R \lambda_{FL} \right\}-\frac{N u}{8 \pi} \left(\sum_{F=1}^n \sigma_F\right)^2.
\label{EffectiveLagrangian}
\end{align}
This effective Lagrangian can be used to calculate mass eigenvalues for the boson and fermion excitations on the vacuum.  Were it not for the last line in (\ref{EffectiveLagrangian}) this would be a very simple matter as all mass terms are already diagonalized.  The last term in (\ref{EffectiveLagrangian}) presents a modification to one particular combination, the field
\begin{equation}
\sigma_u = \frac{1}{\sqrt{n}} \sum_{F=1}^n \sigma_F\, ,
\end{equation}
where $1/\sqrt{n}$ is the normalization factor.  It is a simple algebraic matter to diagonalize the fields $\sigma_F$ to determine the mass eigenvalues.  We find
\begin{equation}
m_{\sigma_u} = 2\Lambda \sqrt{1+\frac{n u}{2}}\,,
\end{equation}
with the $n-1$ remaining boson fields with mass $m=2\Lambda$.

The diagonalization of the fermions is slightly trickier as we must separate the original states with mass terms $\langle \sigma_F \rangle = \Lambda$ from those with $\langle \sigma_F \rangle = -\Lambda$.  For the fermion mass terms we therefore should write
\begin{equation}
\mathcal{L}_{m_f} = 2{\rm i} \Lambda \sum_{f=1}^{n/2}\left[ \tilde{\lambda}_{+fR}\tilde{\lambda}_{+fL}-\tilde{\lambda}_{-fR}\tilde{\lambda}_{-fL}+\sqrt{2u}\zeta_R(\tilde{\lambda}_{+fL}+\tilde{\lambda}_{-fL}) \right].
\end{equation}
Here we have labeled the fermion field $\lambda_{\pm}$ by a subscript indicating the sign of the mass term in the Lagrangian.  We have also indicated the canonically normalized fields
\begin{equation}
\tilde{\lambda}_f = \sqrt{\frac{N}{8\pi}}\lambda_f\,.
\end{equation}

At this point diagonalization is again a matter of algebra, and we find the following mass eigenstates:
\begin{equation}
m\left( \tilde{\lambda}_{-uR} +\sqrt{\frac{nu}{2}}\zeta_R; \tilde{\lambda}_{+uL}\right) = 2\Lambda\sqrt{1+\frac{nu}{2}},
\end{equation}
where
\begin{equation}
\tilde{\lambda}_{-uR} = \frac{1}{\sqrt{n}}\sum_{f=1}^{n/2}\tilde{\lambda}_{-fR}, \mbox{ and }\tilde{\lambda}_{+uL} = \frac{1}{\sqrt{n}}\sum_{f=1}^{n/2}\tilde{\lambda}_{+fL}.
\end{equation}
The orthogonal field combination
\begin{equation}
\sqrt{\frac{nu}{2}}\tilde{\lambda}_{-uR} - \zeta_R
\end{equation}
is a massless right mover which does not interact with the remaining fields in the large-$N$ limit.  All other remaining orthogonal combinations of fermion fields do not get a mass modification, and therefore remain at $m = 2\Lambda$.

Counting the number of boson and fermion states for each value of the mass, we can see that the degrees of freedom match, with an extra sterile massless right-moving fermion field.  Thus, this is another confirmation of our statement that in the large $N$ limit $\mathcal{N} = (1,1)$ supersymmetry is unbroken (provided $n$ is even).

\section{The Witten index}
\label{WI}

The Witten index $I_W$ for the two-dimensional (2,2) $CP(N-1)$ and (1,1) $O(N)$ models have been known since the invention of the index method in \cite{Witten:1982df}, where it was shown that the index is the Euler characteristic of the target manifold:
\begin{equation}
I_{W, CP(N-1)} = N, \mbox{ and } I_{W, O(N)} = 1+(-1)^N.
\end{equation}
For the $O(N)$ model the supersymmetry is unbroken regardless of the value of $N$.  Despite the vanishing of the index for $N$ odd, a modified index can be defined by using the isotopic parity whereby one field $n^i \rightarrow -n^i$.  Defining the operator $K$ as the isotopic parity operator, it can be shown
\begin{equation}
I_K \equiv Tr(-1)^F K = 1-(-1)^N.
\end{equation}
Thus either $I_W$ or $I_K$ are non-vanishing for the $\mathcal{N} = (1,1)$ $O(N)$ model, and supersymmetry is always unbroken \cite{Witten:1982df}.

\vspace{2mm}

The indices for the $O(N)$ model were originally calculated by introducing a height function in the form of a magnetic field on the target space manifold \cite{Witten:1982df}.  The effect of this added potential function is to break the infinite vacuum degeneracy to two distinct vacua.  One may then calculate the fermion mass eigenvalues on each of the vacua to determine the relative fermion number parity between the vacuum states.

For the minimal $\mathcal{N} = (0,1)$ heterotic $O(N)$ model (\ref{MinimalModelComp}), the previous analysis fails since the fermion mass matrix is constrained by the non-existence of right moving $\psi^i_R$ in the Lagrangian.  To calculate the index $I_W$ we consider the case at finite but large $N$.  In this case one may select the vacuum field configuration as
\begin{equation}
\langle n^N\rangle = \pm \frac{1}{g},
\end{equation}
with all other $\langle n^{i \neq N} \rangle = 0$.  The non-zero value of $n^N$ presents a non-trivial mass term in the effective Lagrangian
\begin{equation}
\Delta \mathcal{L} = \pm \sqrt{\frac{4\pi}{N}}\frac{\Lambda}{g} \tilde{\lambda}_R \psi^N_L,
\end{equation}
for canonically normalized $\tilde{\lambda}_R$.  The two vacua present fermi mass terms of opposite sign.  Thus they differ in fermi parity $(-1)^F$, and the Witten index vanishes.  Indeed this is expected from the earlier analysis of the effective potential at $N\to\infty$.  After integration of the $\psi^i_L$ fields only the massless $\lambda_R$ field appears in the Lagrangian.  Two distinct vacua appear with opposite fermion parity due to the massless creation operator $\lambda_R$.  This illustrates the Hoohn-Stolz conjecture that any target space with Riemann metric of positive Ricci curvature has equal numbers of boson and fermion states when right-moving fermions are absent \cite{Stolz:1996} (see also \cite{Yagi:2010tp}).

\vspace{2mm}

Now let us pass to  the non-minimal connected models.  For the moment consider vanishing $u$.  In this case the $\zeta_R$ field is a free Majorana right-moving fermion field with vanishing mass.  Being Majorana, the operator $\zeta_R$ both creates and destroys particles with the same quantum numbers.  Thus, no fermion charge $F$ (not to be confused with the index $F$ of the $O(N)$ factors) can be defined.  However, the fermion parity $(-1)^F$ is well defined (at least in the topologically trivial sector):
\begin{equation}
(-1)^F \zeta_R\ket{0}= -\zeta_R\ket{0}.
\end{equation}

We can see that in the limit of vanishing $u$, any bosonic vacuum $\ket{0}$ of the model is degenerate with the fermionic vacuum $\zeta_R\ket{0}$, and thus the Witten index vanishes:
\begin{equation}
I_W \equiv {\rm Tr}(-1)^F = 0.
\end{equation}

A similar result occurs in the $CP(N-1)^n$ models considered in \cite{Shifman:2014lva}, and just as in that case a modified index can be defined that is non-vanishing for even $n$ and large (and also even) $N$.  There are several arguments to show this result and we will pick the most informative and refer the reader to \cite{Shifman:2014lva} for additional arguments that follow analogously in our case.

We can see that the Lagrangian (\ref{LagrangianComp}) is invariant at the classical level to the chiral parity transformation
\begin{equation}
\psi_L^F \rightarrow \psi_L^F, \;\; \psi_R^F \rightarrow -\psi_R^F, \;\; \zeta_R \rightarrow -\zeta_R.
\label{ChiralParity}
\end{equation}
From the equations of motion we have a chiral condensate for each individual family member $F$
\begin{equation}
\frac{1}{g^2}\langle \psi_L^F \psi_R^F \rangle=\langle \sigma_F \rangle = \pm \Lambda.
\label{ChiralCondensate}
\end{equation}
This one-loop result is exact in the  limit $N\to\infty$.  Thus the chiral parity (\ref{ChiralParity}) is broken at the one loop level due to the chiral condensate.  

There is however a symmetry transformation $P$ preserved at the quantum level that is a combination of the chiral parity (\ref{ChiralParity}) and an exchange of flavours $F$ which we define as follows.  As discussed above the vanishing of the effective potential (\ref{EffectivePotential}) occurs when $\langle \sigma_F \rangle = +\Lambda$ for half of the families and $-\Lambda$ for the remaining half. 
Let us number the sectors with $\langle \sigma_F \rangle = +\Lambda$ as $F=1, 3, ...$,
while the sectors  with $\langle \sigma_F \rangle = -\Lambda$ as $F=2, 4, ...$.
Performing such  the chiral parity transformation (\ref{ChiralParity}) and, simultaneously the shift $F\to F+1$ we see that (\ref{ChiralCondensate}) is invariant.  Following the lessons from \cite{Witten:1982df} we may thus define a modified index $I_P$:
\begin{equation}
I_P \equiv {\rm Tr} (-1)^F P,
\end{equation}
where $P$ is the combineshift in $F$ by one untit and chiral parity transformation.  If the index $I_P$ is non-vanishing then supersymmetry is unbroken.

Defining the creation/annihilation operator of massless $\zeta_R$ states with vanishing energy as $\zeta_{R,0}$ we can write the fermi vacuum states as:
\begin{equation}
\ket{0_F} \equiv \zeta_{R,0}\ket{0_B}, \mbox{ and } P\zeta_{R,0} = -\zeta_{R,0}P.
\end{equation}
Thus, it is trivial to show
\begin{align}
I_P = \bra{0_B}(-1)^F P\ket{0_B}+ \bra{0_F}(-1)^F P\ket{0_F} = 2.
\end{align}
The non-vanishing value of the modified index $I_P$ for even $n$ provides a robust protection of supersymmetry in the large $N$ limit of the connected $O(N)^n$ models in just the same way as in the $CP(N-1)^n$ case.

Additional arguments demonstrating this result can be performed by showing  the vanishing of the order parameter $\langle G \rangle$ at the quantum level.  These arguments can be found in \cite{Shifman:2014lva}.

\section{Conclusions}
We set out to discuss in analogy with the connected $CP(N-1)^n$ model considered in \cite{Shifman:2014lva} the corresponding connected $O(N)^n$ model with the $\mathcal{N}=(0,1)$ preserving deformation generated by a right moving fermion field $\zeta_R$.  The real manifold structure of the $O(N)$ models present some computational differences with the complex $CP(N-1)$ case, however the results on the supersymmetric behavior of the two models are the same.  We have shown that under certain constraints the connected $O(N)^n$ models connected by a right moving fermion preserve the $\mathcal{N} = (0,1)$ supersymmetry.  This was demonstrated for the large $N$ limit by considering the particle spectra and showing the equality between fermion and boson masses. 

We also considered the Witten index $I_W$ of the $O(N)^n$ models, which was shown to vanish.  Despite this result the supersymmetry is unbroken due to the existence of a modified Witten index $I_P$, which can be defined under the exchange symmetry for the case of even $n$.  This is precisely the case observed in the $CP(N-1)^n$ model considered in \cite{Shifman:2014lva}.

For completeness we have also considered the minimal $\mathcal{N} = (0,1)$ heterotic $O(N)$ model which is not excluded by anomalies of the target space manifold.  Our analysis shows that supersymmetry is broken in this model.

\section*{Acknowledgements}
The work of A.P. is supported by the Doctoral Dissertation Fellowship and the Robert O. Pepin Fellowship at the University of Minnesota.  The work of M.S. is supported in part by DOE Grant Number DE-SC0011842.  E.K. is grateful to Vladimir Bychkov for illuminating discussions.

\section*{Appendix: Notation and conventions}

\renewcommand{\theequation}{A.\arabic{equation}}
\setcounter{equation}{0}

The two dimensional gamma matrices can be defined in a Majorana-Weyl basis in two dimensions.  For our purposes we define them as follows:
\begin{equation}
\gamma^0 = \sigma_2, \;\; \gamma^1 = -{\rm i} \sigma_1, \;\; \gamma^5 = \gamma^0\gamma^1 =
\left(\begin{array}{cc}
-1 & 0 \\
0 & 1 \\
\end{array}\right).
\end{equation}
We also define the two index anti-symmetric symbol in two-dimensions
\begin{equation}
\varepsilon_{\alpha\beta} = -{\rm i}(\gamma^0)_{\alpha\beta} =
\left(\begin{array}{cc}
0 & -1 \\
1 & 0 \\
\end{array}\right).
\end{equation}

It will also prove useful to define the right-moving and left-moving coordinates and their corresponding derivatives
\begin{equation}
x_R = x^0-x^1, \;\; x_L = x^0+x^1, \;\; \partial_R = \frac{1}{2}\left(\partial_0 - \partial_1 \right) \;\;  \partial_L = \frac{1}{2}\left(\partial_0 + \partial_1 \right).
\end{equation}

We define the Majorana-Weyl fermion
\begin{equation}
\psi =
\left(\begin{array}{c}
\psi_L \\
\psi_R \\
\end{array}\right), \;\; \bar{\psi} = \psi^T \gamma^0.
\end{equation}

For superfield definitions we will make use of the following conventions for integration over Grassmann  variables $\theta$
\begin{equation}
\int d^2\theta \bar{\theta}\theta \equiv 1.
\end{equation}

Supertransformations of the coordinates may be written as
\begin{equation}
\theta^\alpha \rightarrow \theta^\alpha + \epsilon^\alpha ,\;\; x^\mu \rightarrow x^\mu - {\rm i} \bar{\theta} \gamma^\mu \epsilon.
\end{equation}
We may thus define the following supercharges and superderivatives in differential form
\begin{equation}
Q_\alpha = -{\rm i} \frac{\partial}{\partial\bar{\theta}_\alpha} + (\gamma^\mu \theta)_\alpha \partial_\mu, \;\; \bar{Q}_\alpha = {\rm i} \frac{\partial}{\partial\theta_\alpha} - (\bar{\theta}\gamma^\mu)_\alpha \partial_\mu,
\end{equation}
\begin{equation}
D_\alpha = \frac{\partial}{\partial\bar{\theta}_\alpha} -{\rm i} (\gamma^\mu \theta)_\alpha \partial_\mu, \;\; \bar{D}_\alpha = \frac{\partial}{\partial\theta_\alpha} +{\rm i} (\bar{\theta}\gamma^\mu)_\alpha \partial_\mu.
\end{equation}

These differential operators obey the following anti-commutation relations:
\begin{equation}
\left\{ Q_\alpha, \bar{Q}_\beta \right\} = 2P_\mu(\gamma^\mu)_{\alpha\beta},\;\; \left\{ D_\alpha, \bar{D}_\beta \right\} = 2(\gamma^\mu)_{\alpha\beta}\partial_\mu, \;\; \left\{Q_\alpha,D_\beta\right\} = 0.
\end{equation}

With these conventions for the coordinate transformations we define the $\mathcal{N} = (1,1)$ superfield
\begin{equation}
\Phi(x^\mu,\theta)=\phi + \bar{\theta}\psi + \frac{1}{2}\bar{\theta}\theta F.
\end{equation}
Under the supertransformations
\begin{equation}
\delta \Phi = {\rm i} \bar{\epsilon}Q\Phi,
\end{equation}
the component fields transform as
\begin{equation}
\delta\phi = \bar{\epsilon}\psi, \;\; \delta \psi = -{\rm i} \partial_\mu \phi \gamma^\mu \epsilon + F\epsilon, \;\; \delta F = -{\rm i}\bar{\epsilon}\gamma^\mu \partial_\mu \psi.
\end{equation}

\end{document}